# Tuning the Polar States of Ferroelectric Films via Surface Charges and Flexoelectricity

*Ivan S. Vorotiahin*[1,2], *Eugene A. Eliseev*,[3] *Qian Li*[4],
*Sergei V. Kalinin*[4*], *Yuri A. Genenko*[1†] *and Anna N. Morozovska*[2,‡],

[1]*Institut für Materialwissenschaft, Technische Universität Darmstadt, Jovanka-Bontschits-Str. 2, 64287 Darmstadt, Germany*

[2] *Institute of Physics, National Academy of Sciences of Ukraine, 46, pr. Nauky, 03028 Kyiv, Ukraine*

[3] *Institute for Problems of Materials Science, National Academy of Sciences of Ukraine, Krjijanovskogo 3, 03142 Kyiv, Ukraine*

[4] *The Center for Nanophase Materials Sciences, Oak Ridge National Laboratory, Oak Ridge, TN 37831*

**Abstract**

Using the self-consistent Landau-Ginzburg-Devonshire approach we simulate and analyze the spontaneous formation of the domain structure in thin ferroelectric films covered with the surface screening charge of the specific nature (Bardeen-type surface states). Hence we consider the competition between the screening and the domain formation as alternative ways to reduce the electrostatic energy and reveal unusual peculiarities of distributions of polarization, electric and elastic fields conditioned by the surface screening length and the flexocoupling strength. We have established that the critical thickness of the film and its transition temperature to a paraelectric phase strongly depend on the Bardeen screening length, while the flexocoupling affects the polarization rotation and closure domain structure and induces ribbon-like nano-scale domains in the film depth far from the top open surface. Hence the joint action of the surface screening (originating from e.g. the adsorption of ambient ions or surface states) and flexocoupling may remarkably modify polar and electromechanical properties of thin ferroelectric films.

---

[*] Corresponding author E-mail: sergei2@ornl.gov  (S.V.K.)
[†] Corresponding author  E-mail: genenko@mm.tu-darmstadt.de (Y.A.G.)
[‡] Corresponding author E-mail: anna.n.morozovska@gmail.com (A.N.M.)

## I. INTRODUCTION

Ferroelectric materials remain the object of endless fascination for applied and fundamental science alike. The fundamental aspect of these materials is the presence of surface and interface bound charges due to the discontinuity of the spontaneous polarization. Since the early days of ferroelectricity, these charges were recognized as the key aspect of the physics of ferroelectric surfaces and interfaces. Indeed, if the polarization charge were uncompensated, it would provide bulk-like contributions to the free energy of materials, *i.e.* the corresponding energy would diverge with the system size. These considerations stimulated the search for mechanisms for lowering of this depolarization energy. One such mechanism is formation of ferroelectric domains, extensively analyzed in classical textbooks [1, 2]. The second is charge screening, either internal "bulk" screening by free charges inside a ferroelecric [3, 4] or external "surface" screening by the free charges in the case of the open or electroded ferroelectric film surface. Surprisingly, until now the domain formation and surface screening mechanisms were considered separately, and the competition between these effects almost escaped the attention of scientific community, despite the fact that these processes are intertwined in thin films.

Surface screening of the bound charges is typically provided by the mobile charges adsorbed from the ambience in the case of high or normal humidity [5, 6, 7, 8, 9] or by internal mobile charges of defect nature [10, 11]. In a specific case of the very weak screening, or its artificial absence due to the experimental treatment (cleaned surface in dry atmosphere, ultra-high vacuum or thick dielectric layer at the surface) the screening charges can be localized at surface states caused by the strong band-bending by depolarization field [12, 13, 14, 15, 16, 17]. For both aforementioned cases the screening charges are at least partially free (i.e. mobile) and the spatial distribution of their quasi two-dimensional density is determined by the polarization distribution near the surface.

Due to the long-range nature of the depolarization effects, the incomplete surface screening of ferroelectric polarization strongly influences the domain structure and leads to pronounced effects both near and relatively far from the surface. The incomplete screening strongly affects domain nucleation dynamics, domain shape and period control in thin film under the open-circuit conditions [2, 18], polar properties of the films placed between imperfect "real" electrodes with the finite Tomas-Fermi screening length [19] or separated from the electrodes by ultra-thin dead layers [20] and spatial gaps [21]. The screening deficiency can induce the appearance of the closure domains near free surfaces in ferroelectrics [2, 22, 23], polarization rotation [24], domain wall broadening in both uniaxial and multiaxial ferroelectrics [25, 26], and crossover between different screening regimes of the moving domain wall - surface junctions [27, 28]. Sometimes the screening charges of electrochemical nature can stabilize the single-domain state in the open-circuited thin films [29, 30, 31, 32, 33, 34]. However, more often the multi-domain phase stability region on phase diagrams is between the paraelectric and homogeneous ferroelectric phases. Thus the critical thickness of the size-induced phase transition into a paraelectric phase can vary in a wide range from several lattice constants [35, 36] to tens or hundreds of nanometers [18] or even to micrometers [1, 13], depending on the geometry (e.g. the gap or dead layer thickness), ferroelectric material parameters, temperature, bulk and surface screening charges concentration and mobility.

Note that it is relatively easy to determine the period of the domain structure analytically only in the framework of the simplest Kittel model that considers 180-degree domain stripes with infinitely thin domain walls and does not consider any of screening mechanism at the free surface [37]. In this case the equilibrium period of the domain stripes corresponds to the free energy minimum that consists of the depolarization field energy and the wall surface energy. Polarization gradient, bulk and surface screening make the analytical solution of the problem impossible, and even the numerical solution becomes rather complicated.

Notably, the domain formation offers several possible pathways, including development of classical antiparallel domain arrays and emerging of closure domains. The competition between the two is controlled by the mechanisms for strain accommodation, in turn closely linked to coupling between polarization and strain. This behavior is described by flexoelectric coupling [38] that can lead to unusual changes of the ferroelastic and ferroelectric domain structure, such as interfacial polarization [39], bichirality [40] and non-Ising features [41]. Sometimes, depending on temperature and flexoelectric coupling strength, relative conductivity of the ferroelectric and ferroelastic domain walls becomes at least one order of magnitude higher than in the single-domain regime [42, 43, 44, 45]. The joint action of flexoelectricity and incomplete surface screening facilitates surprisingly versatile changes of the domain structure (including emerging of polarization rotation, closure domains, etc.) near the surfaces of ferroic films [46, 47, 48, 49, 50] and has a noticeable impact on the thermodynamics [51] and kinetics [52] of polarization reversal.

Here for the first time we explore the competition between domain formation and surface screening in a thin ferroelectric film covered with the surface screening charge of specific nature (Bardeen-type surface states). Special attention is paid to the influence of the Bardeen screening length $\Lambda$ [12] and flexoelectric coupling [38, 53, 54] on the film critical thickness, its transition temperature to a paraelectric phase and domain wall structure. Obtained results show that a nontrivial interplay between the surface screening efficiency, stripe domain period, domain wall broadening and closure domains appearance at the open surface occurs to minimize the electro-elastic energy of the ferroelectric film.

## II. PROBLEM STATEMENT AND BASIC EQUATIONS

We consider a ferroelectric film with thickness $h$ placed in a perfect electric contact with conducting bottom electrode that mechanically clamps the film. The top surface of the film is mechanically free and electrically open-circuited, but covered with the surface screening charge due to surface states, or electro-chemically active ions [see **Fig. 1**].

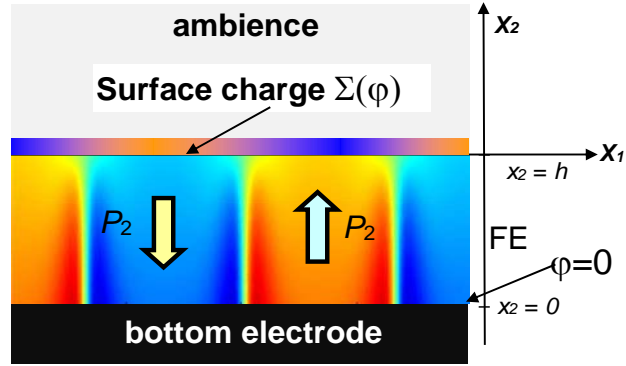

**FIG. 1**. Considered system, consisting of electrically conducting bottom electrode, ferroelectric (FE) film of thickness $h$ with a domain structure (if any exists), surface screening charge with density $\Sigma(\varphi)$ (a model for imperfect screening) and ambient media (from bottom to the top).

The Landau-Ginzburg-Devonshire (**LGD**)-type Gibbs thermodynamic potential of the ferroelectric film is the sum of the bulk ($G_V$) and surface ($G_S$) contributions [44, 48]:

$$G_V = \int_{0<x_2<h} d^3r \left( \begin{array}{c} a_{ik}P_iP_k + a_{ijkl}P_iP_jP_kP_l + a_{ijklmn}P_iP_jP_kP_lP_mP_n + \dfrac{g_{ijkl}}{2}\left(\dfrac{\partial P_i}{\partial x_j}\dfrac{\partial P_k}{\partial x_l}\right) \\ -P_iE_i - \dfrac{\varepsilon_0\varepsilon_b}{2}E_iE_i + Q_{ijkl}\sigma_{ij}P_kP_l - \dfrac{s_{ijkl}}{2}\sigma_{ij}\sigma_{kl} - F_{ijkl}\sigma_{ij}\dfrac{\partial P_l}{\partial x_k} \end{array} \right) - \int_{x_2>h} \dfrac{\varepsilon_0\varepsilon_e}{2}E_iE_i d^3r$$

(1)

The former and the latter integrals represent contributions of a ferroelectric film and an ambient medium, respectively. Summation is performed over all repeating indexes; $P_i$ is a ferroelectric polarization, $E_i = -\partial\varphi/\partial x_i$ is a quasi-static electric field, $\varphi$ is an electric potential. Here we introduced background dielectric permittivity $\varepsilon_b$ [2] and dielectric permittivity of the ambient medium, $\varepsilon_e$. The coefficients of the LGD potential expansion in powers of the polarization are $a_{ik} = \delta_{ik}\alpha_T(T-T_c)$ with the positive constant $\alpha_T$, $a_{ijkl}$ and $a_{ijklmn}$, $T$ is absolute temperature, $T_c$ is the Curie temperature. The elastic stress tensor is $\sigma_{ij}$, $Q_{ijkl}$ is the electrostriction coefficients tensor, $F_{ijkl}$ is the flexoelectric effect tensor [55], $g_{ijkl}$ is the gradient coefficients tensor, $s_{ijkl}$ is the elastic compliances tensor.

The surface energy contains short-range non-electrostatic [56] polarization-dependent contributions from the film surfaces, which have the form

$$G_S = \int_{x_2=0} \left(\frac{\alpha_{S0}}{2} P_i P_j\right) d^2r + \int_{x_2=h} \left(\frac{\alpha_{Sh}}{2} P_i P_j\right) d^2r. \tag{2}$$

The surface energy parameters $\alpha_{S0}$ and $\alpha_{Sh}$ are positive or zero.

The electric potential $\varphi$ obeys the Poisson equation in the film and the Laplace equation in the ambient medium:

$$\varepsilon_0 \varepsilon_b \frac{\partial^2 \varphi}{\partial x_j \partial x_j} = \frac{\partial P_i}{\partial x_i}, \qquad 0 < x_2 < h, \tag{3a}$$

$$\varepsilon_0 \varepsilon_e \frac{\partial^2 \varphi}{\partial x_j \partial x_j} = 0, \qquad x_2 \geq h. \tag{3b}$$

The boundary conditions (**BCs**) for Eqs. (3) assume the vanishing electric potential at the bottom of the film contacting the conducting substrate, and its continuity at the interface between the ferroelectric film and the ambient medium. Another boundary condition at the latter interface requires the equivalence of a discontinuity in the normal component of the electric displacement to the surface free charge:

$$\varphi|_{x_2=0} = 0, \qquad \varphi|_{x_2=h-0} - \varphi|_{x_2=h+0} = 0, \tag{4a}$$

$$\left(P_2 - \varepsilon_0 \varepsilon_b \frac{\partial \varphi}{\partial x_2}\right)\bigg|_{x_2=h-0} + \varepsilon_0 \varepsilon_e \frac{\partial \varphi}{\partial x_2}\bigg|_{x_2=h+0} + \Sigma(\varphi)|_{x_2=h} = 0. \tag{4b}$$

Note that the BC (4b) results directly from the Gauss equation and cannot be obtained from the variation of the Gibbs energy (1)-(2) [57]. Periodic BCs are imposed on the polarization and the electric potential in transverse $x_1$-direction.

Here, we consider the special case of the surface screening charge with the density given by $\Sigma(\varphi) = -\varepsilon_0 \varphi/\Lambda$, where $\Lambda$ is the Bardeen screening length [12, 58]. The period of domain stripes depends on the film thickness *h* and the length $\Lambda$ in a self-consistent way. Besides the Bardeen model [see Fig. 3 and Eq. (17) in Ref. [12]], the expression for $\Sigma(\varphi)$ is relevant for all physical situations assuming that a linear relation between the screening charge density and the electric potential is valid (Tomas-Fermi and Debye-Hückel approximations, physical gaps, etc). For instance, the mathematical form of $\Sigma(\varphi)$ coincides with the Stephenson and Highland model for ionic charge density after a linearization over electric potential at equal ion formation energies [31, 32].

Below we consider a two dimensional (2D) case with polarization components $P_1$ and $P_2$ [see **Fig. 1**], and suppose the cubic symmetry m3m of the parent phase. Minimization of the functional (1) with respect to $P_2$ brings about the Euler-Lagrange equation

$$(2a_1 - 2Q_{12}(\sigma_{11} + \sigma_{33}) - 2Q_{11}\sigma_{22})P_2 - Q_{44}\sigma_{12}P_1 + 4a_{11}P_2^3 + 2a_{12}P_2P_1^2 + 6a_{111}P_2^5 + a_{112}(2P_2P_1^4 + 4P_2^3P_1^2) -$$
$$- g_{11}\frac{\partial^2 P_2}{\partial x_2^2} - g_{44}\frac{\partial^2 P_2}{\partial x_1^2} - (g'_{44} + g_{12})\frac{\partial^2 P_1}{\partial x_1 \partial x_2} + F_{44}\frac{\partial \sigma_{12}}{\partial x_1} + F_{12}\left(\frac{\partial \sigma_{11}}{\partial x_2} + \frac{\partial \sigma_{33}}{\partial x_2}\right) + F_{11}\frac{\partial \sigma_{22}}{\partial x_2} = E_2$$

(5)

Minimization with respect to $P_1$ results in the same form as Eq. (5) with the interchange of subscripts $1 \leftrightarrow 2$. Subscripts 1, 2 and 3, which denote Cartesian coordinates $x_1$, $x_2$, $x_3$ and the Voigt's (matrix) notations are used [59]. BCs for polarization components are the consequence of minimization of the functional (1)-(2):

$$\left(\alpha_{S0}P_2 - g_{11}\frac{\partial P_2}{\partial x_2} - g_{12}\frac{\partial P_1}{\partial x_1} + F_{12}(\sigma_{11} + \sigma_{33}) + F_{11}\sigma_{22}\right)\bigg|_{x_2=0} = 0,$$

$$\left(-\alpha_{Sh}P_2 - g_{11}\frac{\partial P_2}{\partial x_2} - g_{12}\frac{\partial P_1}{\partial x_1} + F_{12}(\sigma_{11} + \sigma_{33}) + F_{11}\sigma_{22}\right)\bigg|_{x_2=h} = 0.$$

(6a)

$$\left(F_{44}\sigma_{12} - g_{44}\frac{\partial P_1}{\partial x_2} - g'_{44}\frac{\partial P_2}{\partial x_1}\right)\bigg|_{x_2=0} = 0, \quad \left(F_{44}\sigma_{12} - g_{44}\frac{\partial P_1}{\partial x_2} - g'_{44}\frac{\partial P_2}{\partial x_1}\right)\bigg|_{x_2=h} = 0 \quad (6b)$$

The generalized fluxes are continuous in $x_1$-direction.

To determine the ***domain period***, we developed the following procedure. We impose periodic boundary conditions on the simulation box along $x_1$ direction, thus closing a bulk on itself. The main problem of this approach is that there can be only even number of domains created in the box, so that a period of simulated domain structures is commensurate with the box size. However, it is still possible to seek for the real size of the domain period, calculating total energy of the system via the energy functional $G$ from Eqs. (1)-(2). In stable states (unlike metastable ones) the system energy is minimal. By varying the box width one can find a width at which the total energy of the given system is minimum while the number of domains is also minimal (i.e. two) unlike any metastable states splitting the structure into several domains or irregularly polarized regions ("band" or "diamond" structures) [see **Fig. S1** in Ref. [60]].

Elastic problem formulation is based on the modified Hooke law obtained using the thermodynamic relation $u_{ij} = -\delta G / \delta \sigma_{kl}$:

$$Q_{ijkl}P_kP_l + F_{ijkl}\frac{\partial P_k}{\partial x_l} + s_{ijkl}\sigma_{kl} = u_{ij}, \quad (7)$$

where $u_{ij}$ are elastic strain tensor components. Mechanical equilibrium conditions $\partial \sigma_{ij}/\partial x_j = 0$ [61] could be rewritten for the considered 2D case as follows:

$$\frac{\partial \sigma_{11}}{\partial x_1} + \frac{\partial \sigma_{12}}{\partial x_2} = 0, \quad \frac{\partial \sigma_{12}}{\partial x_1} + \frac{\partial \sigma_{22}}{\partial x_2} = 0, \quad \frac{\partial \sigma_{31}}{\partial x_1} + \frac{\partial \sigma_{32}}{\partial x_2} = 0. \quad (8)$$

Mechanically free boundary conditions for the mechanical sub-system are imposed at the ferroelectric-outer medium interface ($x_2=h$) and a fixed mechanical displacement $\vec{U}$ is applied at the ferroelectric-substrate interface $\sigma_{12}|_{x_2=h} = 0$, $\sigma_{22}|_{x_2=h} = 0$, $\sigma_{32}|_{x_2=h} = 0$, $U_1|_{x_2=0} = x_1 u_m$, $U_2|_{x_2=0} = 0$, $U_3|_{x_2=0} = 0$. At the box side fictional boundaries we used the conditions $U_2|_{x_1=-w} = U_2|_{x_1=w}$ and $U_1|_{x_1=+w} - U_1|_{x_1=-w} = 2w u_m$ Note that a compressive misfit strain $u_m$ was applied at the film-substrate interface $u_{11} = u_{22} = u_m$ to support vertical direction of polarization (see e.g. Refs. 62, 63, 64).

### III. NUMERICAL RESULTS AND DISCUSSION

The coupled system of Eqs. (3)-(8) along with relevant boundary conditions was analyzed numerically for PbTiO$_3$ (PTO) films. Parameters used in the numerical calculations are listed in **Table I.** We use the so-called "natural case", $\alpha_{S0}=0$ and $\alpha_{Sh}=0$, in numerical calculations, which correspond to the minimal critical thickness of the film [48,49].

**Table I.** Material parameters of ferroelectrics collected and estimated from the Refs

| Parameter of the functional (1)-(2) | Designation and units | Numerical value or interval used in the calculations for ferroelectric PbTiO$_3$ | |
|---|---|---|---|
| background permittivity | $\varepsilon_b$ | 7 | N/A |
| Inverse CW constant | $\alpha^T$ ($\times 10^5$C$^{-2}$·Jm/K) | 3.8 | [65] |
| Curie temperature | $T_C$ (K) | 479+273 | [64, 65] |
| LGD coefficient | $a_{ij}$ ($\times 10^8$C$^{-4}$·m$^5$J) | $a_{11}= -0.73$, $a_{12}= 7.5$ | [64, 61] |
| LGD coefficient | $a_{ijk}$ ($\times 10^8$C$^{-6}$·m$^9$J) | $a_{111}= 2.6$, $a_{112}= 6.1$, $a_{123}= -37.0$ | [64] |
| electrostriction | $Q_{ij}$ (C$^{-2}$·m$^4$) | $Q_{11}=0.089$, $Q_{12}= -0.026$, $Q_{44}=0.0675$ | [66, 61] |
| compliance | $s_{ij}$ ($\times 10^{-12}$ Pa$^{-1}$) | $s_{11}=8.0$, $s_{12}= -2.5$, $s_{44}=9.0$ | [61] |
| gradient coefficients | $g_{ij}$ ($\times 10^{-10}$C$^{-2}$m$^3$J) | $g_{11}=5.1$, $g_{12}= -0.2$, $g_{44}=0.2$ | N/A |
| surface energy coefficient | $\alpha_{Si}$ ($\times 10^{-4}$C$^{-2}$·J) | $\alpha_{S0}=\alpha_{Sh}=0$ | N/A |
| flexoelectric coefficient | $F_{ij}$ ($\times 10^{-11}$C$^{-1}$m$^3$) | $F_{11}= 3$, $F_{12}= 1$, $F_{44}= 5$ | N/A |
| misfit strain | $u_m$ | $-0.01$ | N/A |

To get insight into the interplay between domain formation and surface screening, we analyze the phase diagram of the film in coordinates "temperature $T$ – screening length $\Lambda$" for several values of the film thickness varying in the range (40 – 70) nm [**Fig. 2(a)**]. The transition temperature to the paraelectric phase monotonically decreases with the reduction of the film thickness (compare top and bottom curves). Similarly, the transition temperature monotonically decreases and then saturates with the increase of $\Lambda$. This happens because a very small $\Lambda < 10^{-3}$ nm provides the almost perfect screening. The increase of $\Lambda$ up to 0.1 nm strongly mitigates the screening effectiveness, and $\Lambda \sim 1$ nm or more results in the almost open-circuit boundary conditions at the top surface. Domain stripes with broadened domain walls as well as the appearance of the closure domains near the top surface of the film reduce the depolarization field energy at intermediate $\Lambda$ values but cannot prevent the film transition to a paraelectric phase with the temperature increase and/or $\Lambda$ increase. Therefore the temperature region of paraelectric phase expands with the $\Lambda$ increase and temperature increase, while the multi-domain ferroelectric phase with broadened domain walls (BDW) and closure domains (CD) near the open surface are energetically favorable at smaller temperatures and small and intermediate $\Lambda$ values, respectively. As has been stated by Kittel in Ref. [37], "the domain structure always has its origin in the possibility of lowering the energy of a system by going from a saturated configuration with high energy to a domain configuration with a lower energy". Accordingly, an emergence of the closure domain structure at the top surface of the film is a result of striving for charge neutrality in the absence of a significant compensating surface charge, just as non-emergence of the large closure domains at the bottom is a result of the presence of the bottom electrode which compensates electric charges at the bottom film surface.

We have further obtained that the phase boundary between the paraelectric and ferroelectric phases is slightly dependent on the flexocoupling strength $F_{ij}^* = FA \cdot F_{ij}$, where the realistic coefficients $F_{ij}$ are listed in **Table I**. The amplitude $FA$ was varied in the range from $-1$ to $+1$, since the components $F_{ij}$ were not measured experimentally for PTO, but some components were calculated from the first principles for different perovskites in Refs. [67, 68]. The values in Table I are of the same order as the ones measured for SrTiO$_3$ [69, 70].

Changes in the distribution of the vertical component of the spontaneous polarization, $P_2$, throughout a 60-nm PTO film occurring with the increase of $\Lambda$ values are shown in **Fig. 2(b)-(g)**. At very small $\Lambda$ the screening is almost perfect and the domain wall shape is not perturbed by the top surface [**Fig. 2(b)**]. The small broadening of the domain walls near the top surface appears and increases with $\Lambda$ increase [**Fig. 2(c)-(d)**]. Further increase of $\Lambda$ leads to the formation [**Fig. 2(e)**] and lateral growth [**Fig. 2(f)**] of the closure domains, which eventually merge together at the surface and form an ultra-thin layer with almost zero vertical polarization [**Fig. 2(g)**].

Dependencies of the electric potential, the vertical components of the electric field and polarization at $x_2=h$ on the length $\Lambda$ are shown in **Figs. S2-S3** in Ref. [60]. They are monotonic with saturation at high $\Lambda$. As anticipated, surface electric potential and field tend to zero at $\Lambda \to 0$, the expressions derived in this limit for a homogeneous polarization, $\varphi = \frac{P_2}{\varepsilon_0} \frac{\Lambda}{\varepsilon_b \Lambda + h}$ and $E_2 = -\frac{P_2}{\varepsilon_0} \frac{\Lambda}{\varepsilon_b \Lambda + h}$, are valid subject to the absence of applied voltage. Moreover, we should recognize that, in accordance with our numerical calculations, the energy of a single-domain state in films thicker than 40 nm is very close to the multi-domain ones for $\Lambda < 0.1$ nm. Thus the periodic domain structures shown in **Figs. 2 (b)-(e)** can be either stable or metastable configurations with a very long life-time, at that their period depends on the surface screening length, film thickness and temperature [see **Figs. S4** in Ref. [60]].

Note that the transition temperature (shown in **Fig. 2(a)**) saturates when the closure domains merge together, since the merging creates an ultra-thin layer with very high relative dielectric permittivity acting as an additional screening layer in addition to the surface charge Σ. As a matter of fact, **Fig. 2** illustrates the key result of this work exhibiting how the interplay between the surface screening and domain formation gives rise to complex domain structures (stripe domains, broadened stripes, closure domains, their splitting, etc.).

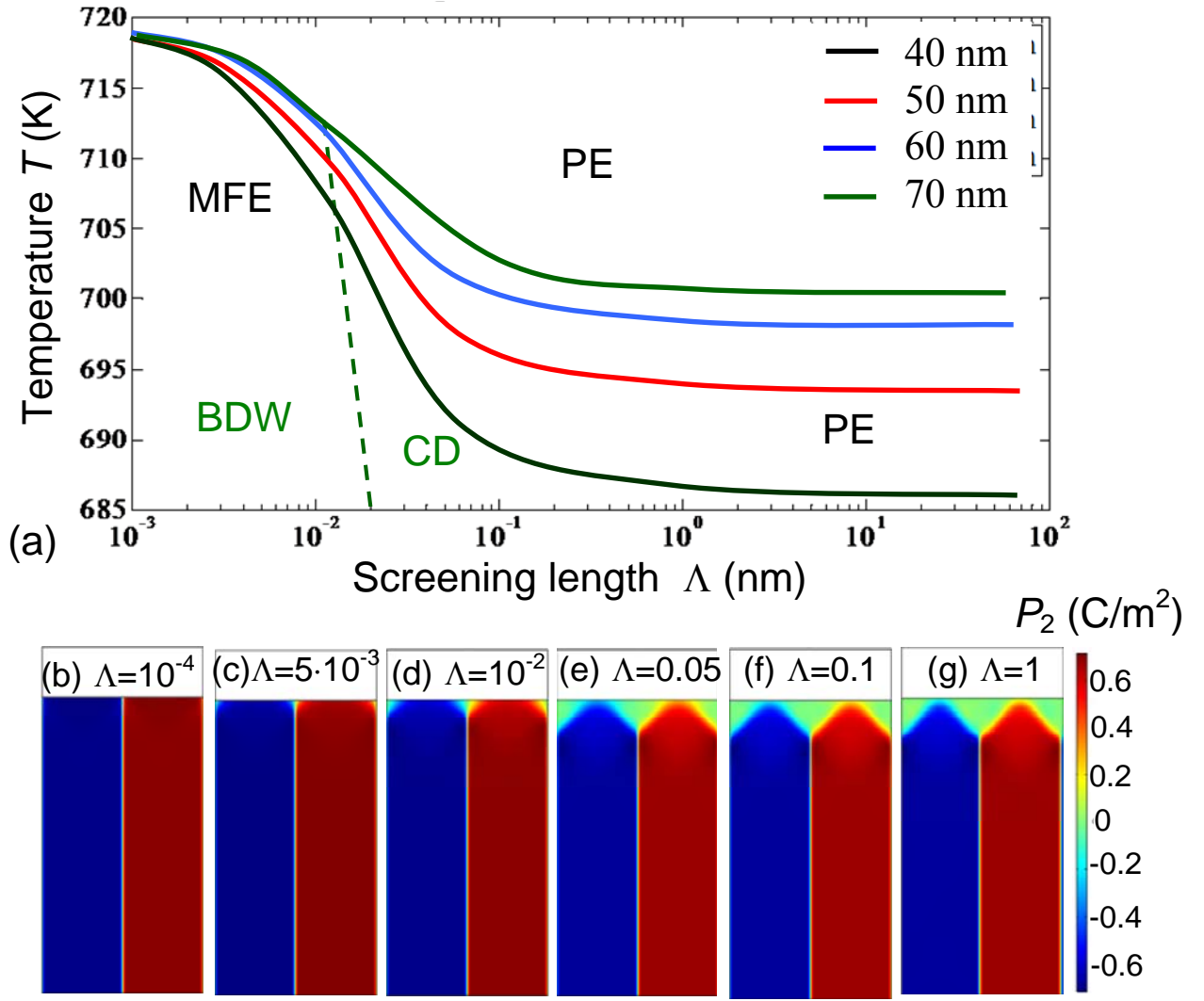

**FIG. 2.** (a) Phase diagrams in coordinates "temperature – screening length" calculated for different film thickness $h$ = 40, 50, 60 and 70 nm (black, red, blue and green curves). There are regions of paraelectric phase (**PE**), multidomain ferroelectric phase (**MFE**) with and/or without broadened domain walls (**BDW**) and closure domains (**CD**) near the top surface of the film. **(b)-(g)** Distributions of the vertical component of the spontaneous polarization $P_2$ throughout a 60-nm PTO film calculated at FA=0 for different values of the screening length Λ: 0.0001 nm **(b)**, 0.005 nm **(c)**, 0.01 nm **(d)**, 0.05 nm **(e)**, 0.1 nm **(f)**, and 1 nm **(g)**.

To get insight into the impact of surface screening on a finite size effect, we further analyze the phase diagram of the film in coordinates "temperature $T$ - film thickness $h$" for the fixed screening length $\Lambda=0.1$ nm and different flexocoupling strengths $F_{ij}^* = FA \times F_{ij}$, where the amplitude $FA$ varies in the range from $-1$ to $+1$ [**Fig. 3(a)**]. The transition temperature could be affected by the flexoelectricity via the changes in the total energy of the system. However, we have obtained that the phase boundary between the paraelectric and multi-domain ferroelectric phases only slightly depends on the absolute value of $FA$ for the films thicker than 20 nm, while the critical thickness $h_{cr}$ of the transition into a paraelectric phase is about 3 nm for $FA=0$ and $h_{cr}=12$ nm for $FA=\pm1$ [compare the solid and dotted curves in **Fig. 3(a)**]. Since the flexoeffect makes the **homogeneous** ferroelectric state less energetically favorable, the transition temperature becomes slightly lower. By creating additional structures (e.g. nanodomains of horizontal polarization at the bottom of films) and deforming polarization profiles (see **Fig. 3 (b-i)**), the flexocoupling increases the total energy of the system, making the zero-point of the temperature-dependent total energy circa 1 K lower for each film thickness. Spatial distributions of the polarization, electric field and elastic stress [shown in the color maps in **Figs. 3(b)-(i)** and in **Figs. S5-S7** in Ref. [60]] slightly yet remarkably depend on the flexocoupling strength in thin films. In particular, the closure domains at the top surface are conditioned by the imperfect screening, while their profile near the bottom electrode is affected by the flexoeffect. It appears that the flexoelectricity creates and stabilizes tiny closure nanodomains near the bottom film-electrode interface, increases the mechanical stresses and wall bending in these areas.

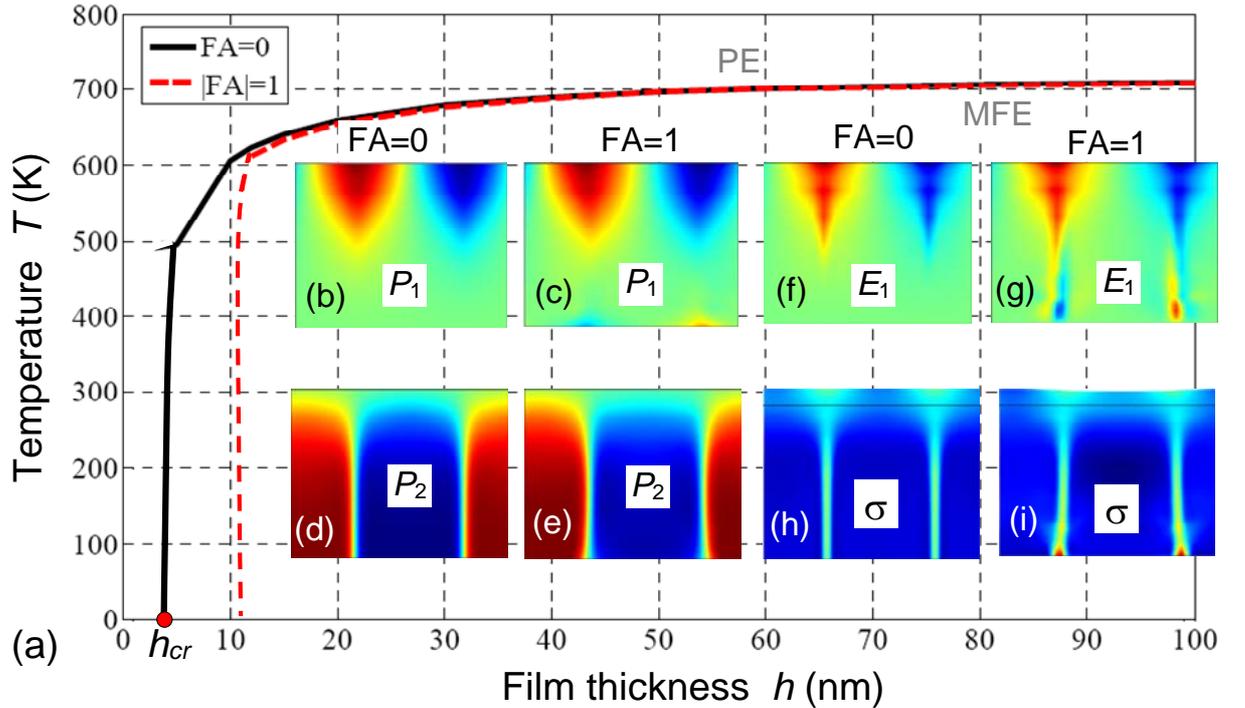

**FIG. 3**. Phase diagrams in coordinates "PTO film thickness – temperature" calculated for $\Lambda=0.1$ nm and flexocoupling amplitudes $FA=0$ (solid curve) and $FA=\pm1$ (dashed curve). There are regions of paraelectric

phase (PE) for thin films and multidomain ferroelectric phase (MFE) with or without closure domains and broadened domain walls near the top surface for thicker films. Spatial distributions of the horizontal $P_1$ **(b, c)** and vertical $P_2$ **(d, e)** polarization components, electric field component $E_1$ **(f, g)** and von Mises elastic stress invariant σ **(h, i)** calculated for film thickness $h$=12 nm, $T$=611 K, Λ=0.1 nm, with ($FA$ = 1) and without ($FA$ = 0) flexocoupling are shown in insets. Notable features near the free top surface and bottom interface with the rigid electrode may be observed.

The domain period appears to be almost independent of the flexocoupling amplitude $FA$, while its dependence on the screening length Λ is a bit stronger with a smooth minima [see **Fig. S4** in Ref. [60]]. Notably, the expected Kittel-Mitsui-Furuichi (**KMF**) relation connecting the period $W$ of the stripe domain structure with infinitely thin walls and the film thickness $h$, $W \sim \sqrt{h}$, appears invalid in our model, that naturally accounts for domain wall broadening near electrically-open surfaces (via the polarization gradient) and closure domains (via polarization rotation). Instead, a weak dependence of the domain period $W$ on the film thickness $h$ is observed, becoming stronger only with reduction of $h$. Other film properties, including screening length Λ or dead layer thickness, also do not show much influence on the equilibrium period of the system. In particular, we carefully checked that the KMF law does not describe our results in the limit $\Lambda \to \infty$ that is the situation most close to the one considered by Mitsui and Furuichi [71]. One of the possible explanations of our result can be a rapid reduction of the electric field under the surface at the distances far below the structure period $W$ values (see **Figs. S2-7** in Ref. [60]), which can breach a KMF-type balance between the domain wall and electrostatic energies, each of which is dependent on the period [2]. Different relation between these energies can cause different dependencies of the period that is a parameter essentially resulting from the energy minimization. In general, the calculated domain structure periods are greater than those predicted by the KMF law (see in **Fig. 4**) and wane differently from KMF behaviour with the decrease of the film thickness, with an enhanced slope at small thicknesses below 40 nm.

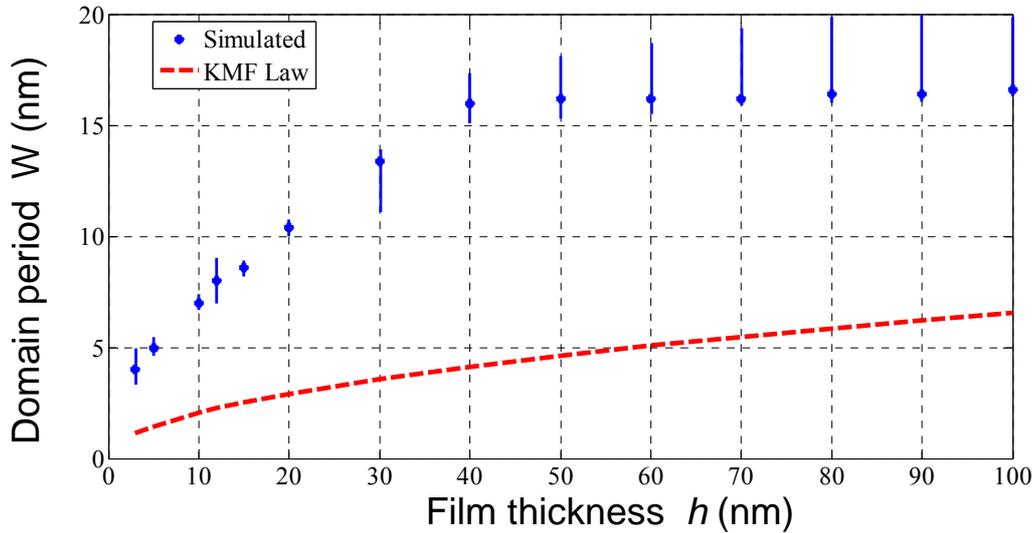

**FIG. 4**. Period of the domain structure W vs. the thickness $h$ of PTO film calculated for considered screening model (points with error bars) and the Kittel-Mitsui-Furuichi law (dashed curve). The data simulated at $\Lambda$=0.1 nm exhibit saturation at around 17 nm at big thicknesses (40-100 nm) and thus does not fit into the KMF law, defined as $W = 2\sqrt{h_M h}$ where $h_M$ is a characteristic length equal to ≈ 0.11 nm in the PTO case.

Finally, we note that though the flexoelectricity affects the system energy and phase diagram relatively weakly it can cause unusual features in the domain morphology. In particular, we have found several cases when the flexocoupling impact facilitates interesting formations in the depth of a thicker film. Exemplarily, **Fig. 5** shows the appearance of ribbon-like nanodomains with domain dimensions (≈10×20 nm) in the depth of the film caused and stabilized by the flexocoupling. The nanodomains are reminiscent of polar nanoregions known in ferroelectric relaxors [72]. Ribbon-like domains far from the top open surface (rather closer to the bottom electroded surface) are critically conditioned by the flexoeffect [**Figs. 5(a)-(b)**], they are absent at FA=0 [**Figs. 5(c)-(d)**]. The film thickness $h$=110 nm is chosen to show the difference of the distributions near the top and bottom surfaces from those in the central part of the film.

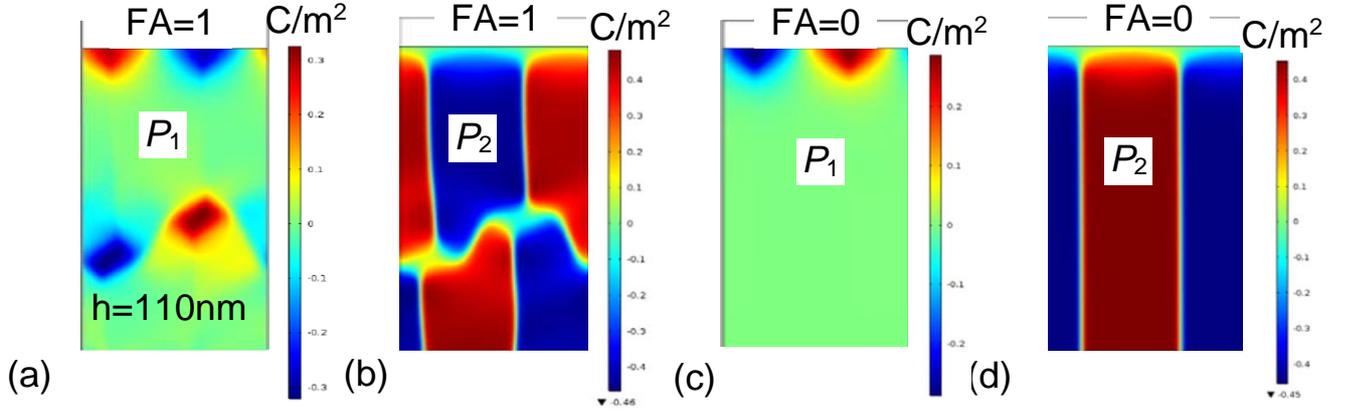

**FIG. 5**. Plots **(a-d)** show spatial distributions of the lateral $P_1$ **(a, c)** and vertical $P_2$ **(b, d)** components of the spontaneous polarization in the 110 nm thick PTO film calculated for Λ=0.1 nm, positive, FA=1 **(a,b),** and zero, FA=0 **(c,d),** flexocoupling amplitudes.

The physical origin of the ribbon-like nanodomains is a metastability dependent on initial and external conditions. The increase of the flexocoupling strength enhances the probability of appearance of unusual inhomogeneous structures. Note that the emergence of the metastable ribbon-like nanodomains does not affect significantly the domain period that remains close to its equilibrium value. It is particularly important that a film in which they form is thick enough to contain such structures deep in the bulk, so that they cannot reach any of the surfaces and "anihilate". Thus, in relatively thick ferroelectric films, the regions with horizontal polarization component can emerge deep in the bulk, enlarging energy of the system and creating "ribbon" or "diamond" structures that split vertical domains in parts. During the numerical calculations, along with minimum-energy two domain states [see the note on the domain period definition] also other, less energetically favourable multidomain states, occur. Typically, such states exhibit few (usually 2 or 3) pairs of narrow domains, however they might form structures with regions of ribbon or diamond shapes as well. Such metastable states have different sources. One of them is numerical and means that the box width is incommensurate with the domain period. By choosing a correct box size these states can be avoided. Another reason can be initial conditions, when a structure is formed that can only relax into a long-living metastable state other than the stable one, although the latter case is extremely rare. Metastable states can emerge only occasionally and rarely if all the requirements for energy minimization are met. However, after introducing the flexocoupling into the system, the metastability becomes much more often, though not entailing significant changes in the period of the domain structure.

**IV. DISCUSSION AND CONCLUSION**

Using a self-consistent Landau-Ginzburg-Devonshire approach we simulated the formation of the domain structure in thin ferroelectric films covered with the surface screening charge of the specific nature (Bardeen-type surface states) and analyzed unusual features of the polarization distribution, the electric and elastic fields conditioned by the surface screening length and the flexocoupling strength.

Paying special attention to the flexoelectric coupling we explored the competition between the domain formation and the surface screening, as alternative ways to reduce the electrostatic energy, and constructed the phase diagrams showing dominance of the domain splitting for weak screening. We established that the film critical thickness and its transition temperature to a paraelectric phase strongly depend on the Bardeen screening length and very weakly depend on the flexocoupling, while the flexocoupling influences the polarization rotation and closure domain structure. Particularly, the screening length increase leads to the essential reduction of the transition temperature to the paraelectric phase. Flexoelectric coupling also leads to a slight decrease of the transition temperature, has a small yet remarkable effect on the domain structure near the film surfaces and causes nanoscopic closure domains at the rigid contact with the bottom conducting electrode.

Surprisingly, ribbon-like nanodomains occasionally emerge due to the flexoeffect also in the film depth far from the surfaces. This principal observation might be related to a still highly disputable formation mechanism of the so-called polar nanoregions in relaxor ferroelectrics. Being exotic metastable formations boosted by flexoelectricity in the conventional ferroelectric PTO such structures could be stabilized in relaxor ferroelectrics by chemical or structural disorder.

Dependence of the period of the stripe domain structure on the screening length and on the flexoelectric coupling appears to be weak. Its monotonically increasing dependence on the film thickness is, in contrast, pronounced but clearly distinct from the classical Kittel-Mitsui-Furuichi square root law. The latter disagreement must not be surprising since the classical model neglects both the structure of the domain walls and the formation of the closure domains near the free surface. Nevertheless, a virtually constant domain width for large film thicknesses above some characteristic value (in our simulations, above $h$~40 nm) could hardly be expected. This behaviour might disclose the fact that the domain wall energy in thick films is not proportional to the domain wall length.

**ACKNOWLEDGEMENTS**

I.S.V. gratefully acknowledges support from the Deutsche Forschungsgemeinschaft (DFG) through the grant GE 1171/7-1. E.A.E. and A.N.M. acknowledge the Center for Nanophase Materials Sciences, which is a DOE Office of Science User Facility, CNMS2016-061. S.V.K. research is sponsored by the Division of Materials Sciences and Engineering, BES, US DOE. A portion of this research (S.V.K.) was conducted at the Center for Nanophase Materials Sciences, which is a DOE Office of Science User Facility.

# Supplementary Materials to the manuscript
# "Tuning the Polar States of Ferroelectric Films via Surface Charges and Flexoelectricity"

Ivan S. Vorotiahin[1,2], Eugene A. Eliseev,[3] Qian Li[4],

Sergei V. Kalinin[4*], Yuri A. Genenko[1†] and Anna N. Morozovska[2,‡],

[1]Institut für Materialwissenschaft, Technische Universität Darmstadt, Jovanka-Bontschits-Str. 2,

64287 Darmstadt, Germany

[2] Institute of Physics, National Academy of Sciences of Ukraine,

46, pr. Nauky, 03028 Kyiv, Ukraine

[3] Institute for Problems of Materials Science, National Academy of Sciences of Ukraine,

Krjijanovskogo 3, 03142 Kyiv, Ukraine

[4] The Center for Nanophase Materials Sciences, Oak Ridge National Laboratory,

Oak Ridge, TN 37831

---

[*] Corresponding author E-mail: sergei2@ornl.gov   (S.V.K.)
[†] Corresponding author    E-mail: genenko@mm.tu-darmstadt.de (Y.A.G.)
[‡] Corresponding author E-mail: anna.n.morozovska@gmail.com (A.N.M.)

# APPENDIX A.

Here we consider only 2D case with two polarization components. For the sake of clarity let us denote "$x_1$" as "x", "$x_2$" as "y", and "$x_3$" as "z", therefore one could reduce free energy (1) to the following form (see e.g. [i]). The bulk contribution is ($G_V$)

$$G_V = \int_{0<y<h} d^3r \begin{pmatrix} a_1(P_x^2 + P_y^2) + a_{11}(P_x^4 + P_y^4) + a_{12}P_x^2 P_y^2 + a_{111}(P_x^6 + P_y^6) + a_{112}(P_x^2 P_y^4 + P_x^4 P_y^2) \\ \frac{g_{11}}{2}\left(\left(\frac{\partial P_x}{\partial x}\right)^2 + \left(\frac{\partial P_y}{\partial y}\right)^2\right) + \frac{g_{44}}{2}\left(\left(\frac{\partial P_x}{\partial y}\right)^2 + \left(\frac{\partial P_y}{\partial x}\right)^2\right) + g'_{44}\frac{\partial P_x}{\partial y}\frac{\partial P_y}{\partial x} + g_{12}\frac{\partial P_x}{\partial x}\frac{\partial P_y}{\partial y} \\ -P_x E_x - P_y E_y - \frac{\varepsilon_0 \varepsilon_b}{2}\left(E_x^2 + E_y^2\right) \\ -\left(Q_{12}(\sigma_{yy} + \sigma_{zz}) + Q_{11}\sigma_{xx}\right)P_x^2 - \left(Q_{12}(\sigma_{xx} + \sigma_{zz}) + Q_{11}\sigma_{yy}\right)P_y^2 - Q_{44}\sigma_{xy}P_x P_y \\ -\left(F_{44}\sigma_{xy}\frac{\partial}{\partial y} + F_{12}\left(\sigma_{yy}\frac{\partial}{\partial x} + \sigma_{zz}\frac{\partial}{\partial x}\right) + F_{11}\sigma_{xx}\frac{\partial}{\partial x}\right)P_x \\ -\left(F_{44}\sigma_{xy}\frac{\partial}{\partial x} + F_{12}\left(\sigma_{xx}\frac{\partial}{\partial y} + \sigma_{zz}\frac{\partial}{\partial y}\right) + F_{11}\sigma_{yy}\frac{\partial}{\partial y}\right)P_y \\ -\frac{s_{11}}{2}\left(\sigma_{xx}^2 + \sigma_{yy}^2 + \sigma_{zz}^2\right) - s_{12}\left(\sigma_{xx}\sigma_{yy} + \sigma_{yy}\sigma_{zz} + \sigma_{xx}\sigma_{zz}\right) - \frac{s_{44}}{2}\left(\sigma_{xy}^2 + \sigma_{yz}^2 + \sigma_{xz}^2\right) \end{pmatrix}$$

$$- \int_{y>h} d^3r \frac{\varepsilon_0 \varepsilon_e}{2}\left(E_x^2 + E_y^2\right)$$

(S1.1a)

The first and the last integrals represent the contributions of ferroelectric film and external media, respectively. The surface contribution ($G_S$) has the following form

$$G_S = \int_{x_2=h} \left(\frac{\alpha_{Sx}}{2}P_x^2 + \frac{\alpha_{Sy}}{2}P_y^2\right) d^2r \qquad (S1.1b)$$

Surface contribution (S1.1b) contains only polarization dependent term. $P_i$ is a ferroelectric polarization, $E_i = -\partial\varphi/\partial x_i$ is a quasi-static electric field, φ is the electric potential. Here we introduced background dielectric permittivity $\varepsilon_b$ and dielectric permittivity of outer media, $\varepsilon_e$. The coefficients of LGD potential expansion on the polarization powers are $a_{ik} = \delta_{ik}\alpha_T(T-T_c)$, $a_{ijkl}$ and $a_{ijklmn}$, $T$ is the absolute temperature, $T_c$ is the Curie temperature. This choice of LGD expansion corresponds to materials with inversion center in the parent phase (e.g. with cubic parent phase). Elastic stress tensor is $\sigma_{ij}$, $Q_{ijkl}$ is electrostriction tensor, $F_{ijkl}$ is the flexoelectric effect tensor [ii], $g_{ijkl}$ is gradient coefficient tensor, $s_{ijkl}$ is elastic compliance tensor.

Subscripts 1, 2 and 3 denote Cartesian coordinates *x*, *y*, *z* and Voigt's (matrix) notations are used:

$$a_{11} \equiv a_1, \quad a_{1111} \equiv a_{11}, \quad 6a_{1122} \equiv a_{12}, \qquad (S1.2a)$$

$$g_{1111} \equiv g_{11}, \quad g_{1122} \equiv g_{12}, \qquad (S1.2b)$$

$$g_{1212} \equiv g_{44}, \quad g'_{44} = g_{1221}, \quad \text{(at that it is possible } g_{44} \neq g'_{44}\text{)}, \qquad (S1.2c)$$

$$Q_{1111} \equiv Q_{11}, \quad Q_{1122} \equiv Q_{12}, \quad 4Q_{1212} \equiv Q_{44}, \qquad (S1.2d)$$

$$s_{1111} \equiv s_{11}, \quad s_{1122} \equiv s_{12}, \quad 4s_{1212} \equiv s_{44}, \qquad (S1.2e)$$

$$F_{1111} \equiv F_{11}, \quad F_{1122} \equiv F_{12}, \quad 2F_{1212} \equiv F_{44}. \qquad (S1.2f)$$

Note that different factors (either "4", "2" or "1") in the definition of matrix notations with indices "44" are determined by the internal symmetry of tensors as well as by the symmetry of the corresponding physical properties tensors (see e.g. [iii]). Here we suppose cubic symmetry m3m of the parent phase.

Let us consider a ferroelectric film with thickness *h* on the rigid conductive substrate. The minimization of free energy (S1.1) with respect to the electric potential φ gives the Poisson equation

$$\varepsilon_0 \varepsilon_b \frac{\partial^2 \varphi}{\partial x_j \partial x_j} = \frac{\partial P_i}{\partial x_i} \quad 0 < x_2 < h \qquad (S1.3a)$$

$$\varepsilon_0 \varepsilon_e \frac{\partial^2 \varphi}{\partial x_j \partial x_j} = 0 \quad x_2 \geq h \qquad (S1.3b)$$

Here the summation is performed over all repeating indexes. The following conditions should be met at the interface between ferroelectric and outer media:

$$\left(-\varepsilon_0 \varepsilon_e \frac{\partial \varphi}{\partial x_2}\right)\bigg|_{x_2=h+0} - \left(-\varepsilon_0 \varepsilon_b \frac{\partial \varphi}{\partial x_2} + P_2\right)\bigg|_{x_2=h-0} = \Sigma \qquad (S1.4a)$$

$$\varphi|_{x_2=h-0} - \varphi|_{x_2=h+0} = 0 \qquad (S1.4b)$$

Here we introduced surface screening charge with density

$$\Sigma = -\varepsilon_0 \varphi / \Lambda. \qquad (S1.4c)$$

where Λ is the surface screening length [iv]. The boundary condition at the conducting substrate corresponds to the fixed potential, namely:

$$\varphi|_{x_2=0} = 0 \qquad (S1.5)$$

Since in numerical computations most of the calculations are restricted to finite "calculation domain", we have to consider "fictious boundaries" $x_2 = \pm w$, at which additional boundary conditions should be applied. Periodic boundary conditions for polarization, potential and mechanics are set to the side boundaries, with $A_{src} = A_{dst}$, where $A$ is a correspondent value of the aforementioned physical quantities ($P_1$, $P_2$, V, $\sigma_{ijkl}$) on "source" and "destination" boundaries.

The Euler-Lagrange equations of LGD theory for polarization components $P_x$ and $P_y$ of multiaxial ferroelectric can be obtained from Eq. (S1.1a) as follows

$$(2a_1 - 2Q_{12}(\sigma_{yy} + \sigma_{zz}) - 2Q_{11}\sigma_{xx})P_x - Q_{44}\sigma_{xy}P_y + 4a_{11}P_x^3 + 2a_{12}P_y P_x^2 + 6a_{111}P_x^5 + a_{112}(2P_x P_y^4 + 4P_x^3 P_y^2) -$$
$$- g_{11}\frac{\partial^2 P_x}{\partial x^2} - g_{44}\frac{\partial^2 P_x}{\partial y^2} - (g'_{44} + g_{12})\frac{\partial^2 P_y}{\partial y \partial x} + F_{44}\frac{\partial \sigma_{xy}}{\partial y} + F_{12}\left(\frac{\partial \sigma_{yy}}{\partial x} + \frac{\partial \sigma_{zz}}{\partial x}\right) + F_{11}\frac{\partial \sigma_{xx}}{\partial x} = E_x$$

(S1.6a)

$$(2a_1 - 2Q_{12}(\sigma_{xx} + \sigma_{zz}) - 2Q_{11}\sigma_{yy})P_y - Q_{44}\sigma_{xy}P_x + 4a_{11}P_y^3 + 2a_{12}P_y P_x^2 + 6a_{111}P_y^5 + a_{112}(2P_y P_x^4 + 4P_y^3 P_x^2) -$$
$$- g_{11}\frac{\partial^2 P_y}{\partial y^2} - g_{44}\frac{\partial^2 P_y}{\partial x^2} - (g'_{44} + g_{12})\frac{\partial^2 P_x}{\partial y \partial x} + F_{44}\frac{\partial \sigma_{xy}}{\partial x} + F_{12}\left(\frac{\partial \sigma_{xx}}{\partial y} + \frac{\partial \sigma_{zz}}{\partial y}\right) + F_{11}\frac{\partial \sigma_{yy}}{\partial y} = E_y$$

(S1.6b)

Boundary conditions for polarization are the consequence of the functional (S1.1a) minimization:

$$\left(-g_{44}\frac{\partial P_x}{\partial y} - g'_{44}\frac{\partial P_y}{\partial x} + F_{44}\sigma_{xy}\right)\bigg|_{y=0} = 0, \quad \left(+g_{44}\frac{\partial P_x}{\partial y} + g'_{44}\frac{\partial P_y}{\partial x} - F_{44}\sigma_{xy}\right)\bigg|_{y=h} = 0 \quad \text{(S1.7a)}$$

$$\left(+g_{11}\frac{\partial P_x}{\partial x} + g_{12}\frac{\partial P_y}{\partial y} - F_{12}(\sigma_{yy} + \sigma_{zz}) - F_{11}\sigma_{xy}\right)\bigg|_{x=w} = 0,$$
$$\left(-g_{11}\frac{\partial P_x}{\partial x} - g_{12}\frac{\partial P_y}{\partial y} + F_{12}(\sigma_{yy} + \sigma_{zz}) + F_{11}\sigma_{xx}\right)\bigg|_{x=-w} = 0.$$

(S1.7b)

$$\left(\alpha_{S0}P_y - g_{11}\frac{\partial P_y}{\partial y} - g_{12}\frac{\partial P_x}{\partial x} + F_{12}(\sigma_{xx} + \sigma_{zz}) + F_{11}\sigma_{yy}\right)\bigg|_{y=0} = 0,$$
$$\left(\alpha_{Sh}P_y + g_{11}\frac{\partial P_y}{\partial y} + g_{12}\frac{\partial P_x}{\partial x} - F_{12}(\sigma_{xx} + \sigma_{zz}) - F_{11}\sigma_{yy}\right)\bigg|_{y=h} = 0$$

(S1.7c)

$$\left(-g_{44}\frac{\partial P_y}{\partial x} - g'_{44}\frac{\partial P_x}{\partial y} + F_{44}\sigma_{xy}\right)\bigg|_{x=-w} = 0, \quad \left(+g_{44}\frac{\partial P_y}{\partial x} + g'_{44}\frac{\partial P_x}{\partial y} - F_{44}\sigma_{xy}\right)\bigg|_{x=w} = 0. \quad \text{(S1.7d)}$$

It is seen that one can introduce the generalized polarization flux matrix $\hat{\Gamma}$ with following components

$$\Gamma_{xx} = -g_{11}\frac{\partial P_x}{\partial x} - g_{12}\frac{\partial P_y}{\partial y} + F_{12}(\sigma_{yy} + \sigma_{zz}) + F_{11}\sigma_{xx}, \quad \Gamma_{yx} = -g_{44}\frac{\partial P_x}{\partial y} - g'_{44}\frac{\partial P_y}{\partial x} + F_{44}\sigma_{xy} \quad \text{S1.8a}$$

$$\Gamma_{xy} = -g_{44}\frac{\partial P_y}{\partial x} - g'_{44}\frac{\partial P_x}{\partial y} + F_{44}\sigma_{xy}, \quad \Gamma_{yy} = -g_{11}\frac{\partial P_y}{\partial y} - g_{12}\frac{\partial P_x}{\partial x} + F_{12}(\sigma_{xx} + \sigma_{zz}) + F_{11}\sigma_{yy} \quad \text{(S1.8b)}$$

With this designation the LGD equations (S1.6) can be rewritten as

$$(2a_1 - 2Q_{12}(\sigma_{yy} + \sigma_{zz}) - 2Q_{11}\sigma_{xx})P_x - Q_{44}\sigma_{xy}P_y + 4a_{11}P_x^3 + 2a_{12}P_y^2 P_x + 6a_{111}P_x^5 + a_{112}(2P_x P_y^4 + 4P_x^3 P_y^2)$$
$$+ \frac{\partial \Gamma_{xx}}{\partial x} + \frac{\partial \Gamma_{yx}}{\partial y} - E_x = 0$$
(S1.9a)

$$(2a_1 - 2Q_{12}(\sigma_{xx} + \sigma_{zz}) - 2Q_{11}\sigma_{yy})P_y - Q_{44}\sigma_{xy}P_x + 4a_{11}P_y^3 + 2a_{12}P_y P_x^2 + 6a_{111}P_y^5 + a_{112}(2P_y P_x^4 + 4P_y^3 P_x^2)$$
$$+ \frac{\partial \Gamma_{xy}}{\partial x} + \frac{\partial \Gamma_{yy}}{\partial y} - E_y = 0$$
(S1.9b)

along with the boundary conditions (S1.7) in the form

$$(\alpha_{S0}P_y + \Gamma_{yy})\big|_{y=0} = 0, \quad (\alpha_{Sh}P_y - \Gamma_{yy})\big|_{y=h} = 0. \quad \text{(S1.10a)}$$

$$\Gamma_{xy}\big|_{x=-w} = 0, \quad -\Gamma_{xy}\big|_{x=w} = 0. \quad \text{(S1.10b)}$$

$$\Gamma_{yx}\big|_{y=0} = 0, \quad -\Gamma_{yx}\big|_{y=h} = 0. \quad \text{(S1.10c)}$$

$$\Gamma_{xx}\big|_{x=-w} = 0, \quad -\Gamma_{xx}\big|_{x=w} = 0. \quad \text{(S1.10d)}$$

Elastic subproblem formulation is based on the modified Hooke law in the following form, which can be obtained using the thermodynamic relation $u_{ij} = -\delta G_V / \delta \sigma_{kl}$:

$$Q_{ijkl}P_k P_l + F_{ijkl}\frac{\partial P_k}{\partial x_l} + s_{ijkl}\sigma_{kl} = u_{ij} \quad \text{(S1.11)}$$

where $u_{ij}$ are elastic strain tensor components. Mechanical equilibrium conditions are[v]

$$\frac{\partial \sigma_{ij}}{\partial x_j} = 0 \quad \text{(S1.12a)}$$

which can be rewritten for the considered 2D case as follows:

$$\frac{\partial \sigma_{xx}}{\partial x} + \frac{\partial \sigma_{xy}}{\partial y} = 0, \quad \frac{\partial \sigma_{yx}}{\partial x} + \frac{\partial \sigma_{yy}}{\partial y} = 0, \quad \frac{\partial \sigma_{zx}}{\partial x} + \frac{\partial \sigma_{zy}}{\partial y} = 0 \quad \text{(S1.12b)}$$

Boundary conditions for the mechanical part are the following. At the mechanically free interface ferroelectric / outer media one has

$$\sigma_{xy}\big|_{y=h} = 0, \quad \sigma_{yy}\big|_{y=h} = 0, \quad \sigma_{zy}\big|_{y=h} = 0 \tag{S1.13}$$

We suppose that mechanical displacement $\vec{U}$ is fixed at the interface ferroelectric/substrate

$$U_x\big|_{y=0} = x u_m, \quad U_y\big|_{y=0} = 0, \quad U_z\big|_{y=0} = 0. \tag{S1.14}$$

At the side (fictions) boundaries one could use periodic condition, modified with respect to misfit strain:

$$U_x\big|_{x=+w} - U_x\big|_{x=-w} = 2w u_m, \quad U_y\big|_{x=+w} - U_y\big|_{x=-w} = 0 \tag{S1.15}$$

Taking into account Eqs. (S1.2), Eqs. (S1.11) for cubic symmetry ferroelectrics with the components of polarization $P_x$, $P_y$ can be rewritten as:

$$u_{xx} = s_{11}\sigma_{xx} + s_{12}(\sigma_{yy} + \sigma_{zz}) + Q_{11}P_x^2 + Q_{12}P_y^2 + F_{11}\frac{\partial P_x}{\partial x} + F_{12}\frac{\partial P_y}{\partial y}, \tag{S1.16a}$$

$$u_{yy} = s_{11}\sigma_{yy} + s_{12}(\sigma_{xx} + \sigma_{zz}) + Q_{12}P_x^2 + Q_{11}P_y^2 + F_{11}\frac{\partial P_y}{\partial y}, \tag{S1.16b}$$

$$u_{zz} = s_{11}\sigma_{zz} + s_{12}(\sigma_{yy} + \sigma_{xx}) + Q_{12}P_x^2 + Q_{12}P_y^2 + F_{12}\frac{\partial P_y}{\partial y} \tag{S1.16c}$$

$$u_{xy} = \frac{s_{44}}{2}\sigma_{xy} + \frac{Q_{44}}{2}P_x P_y + \frac{F_{44}}{2}\frac{\partial P_y}{\partial x} + \frac{F_{44}}{2}\frac{\partial P_x}{\partial y} \tag{S1.16d}$$

$$u_{xz} = \frac{s_{44}}{2}\sigma_{xz} \tag{S1.16e}$$

$$u_{yz} = \frac{s_{44}}{2}\sigma_{yz} \tag{S1.16f}$$

Note that denominator "2" appearance in Eqs. (S1.16d)-(S1.16f) is related to the fact, that we use tensor notation for strain and stress components, and matrix notations for compliances. Parameters to be used are listed in Table 1.

**APPENDIX B**

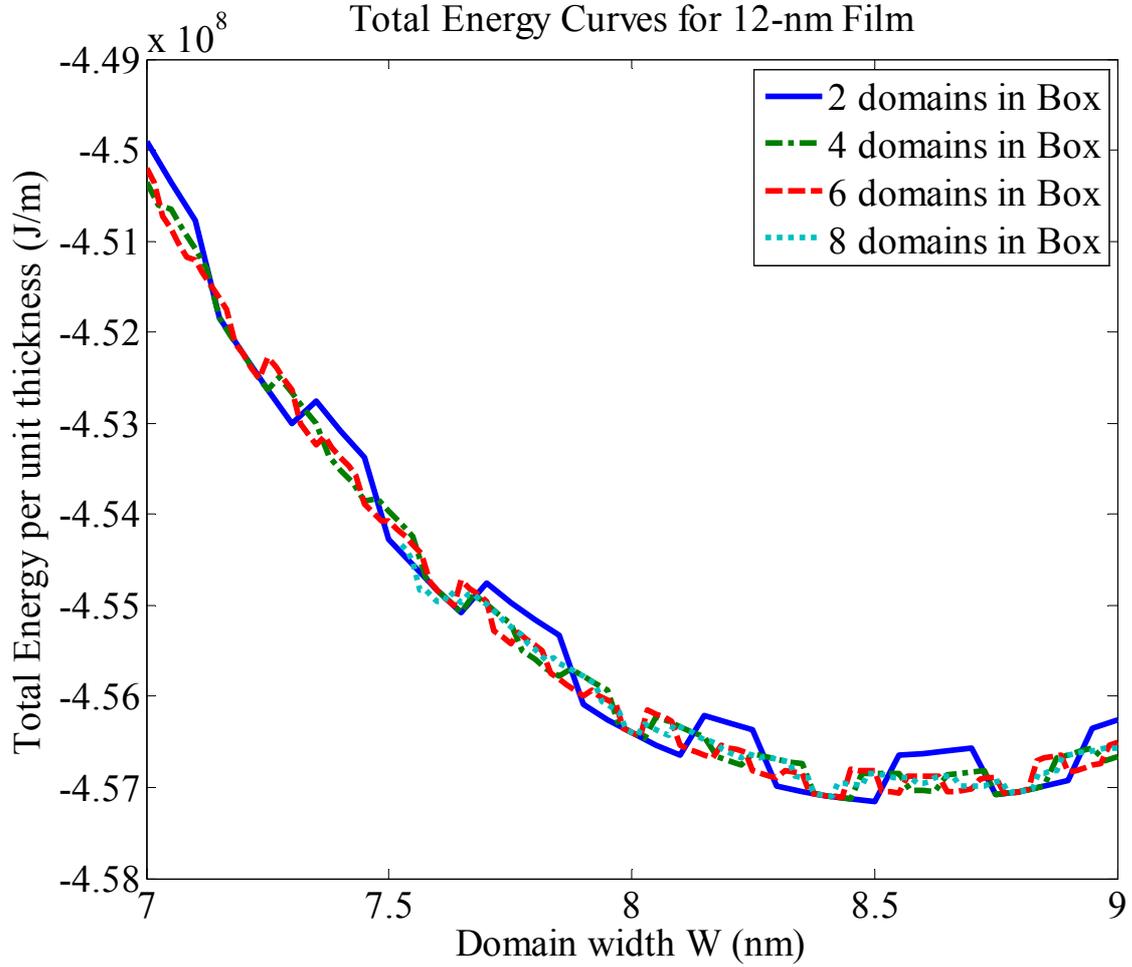

**FIG. S1**. Evaluation of the equilibrium domain width for 12-nm film through the minimum of total energy of the system at room temperature and $\Lambda=0.1$ nm. Different curves correspond to the different size orders of the simulation box that can contain 2, 4, 6, and 8 domains, having minimized energy. It can be seen that the domain width with minimum energy for this film is estimated as ≈8.5.

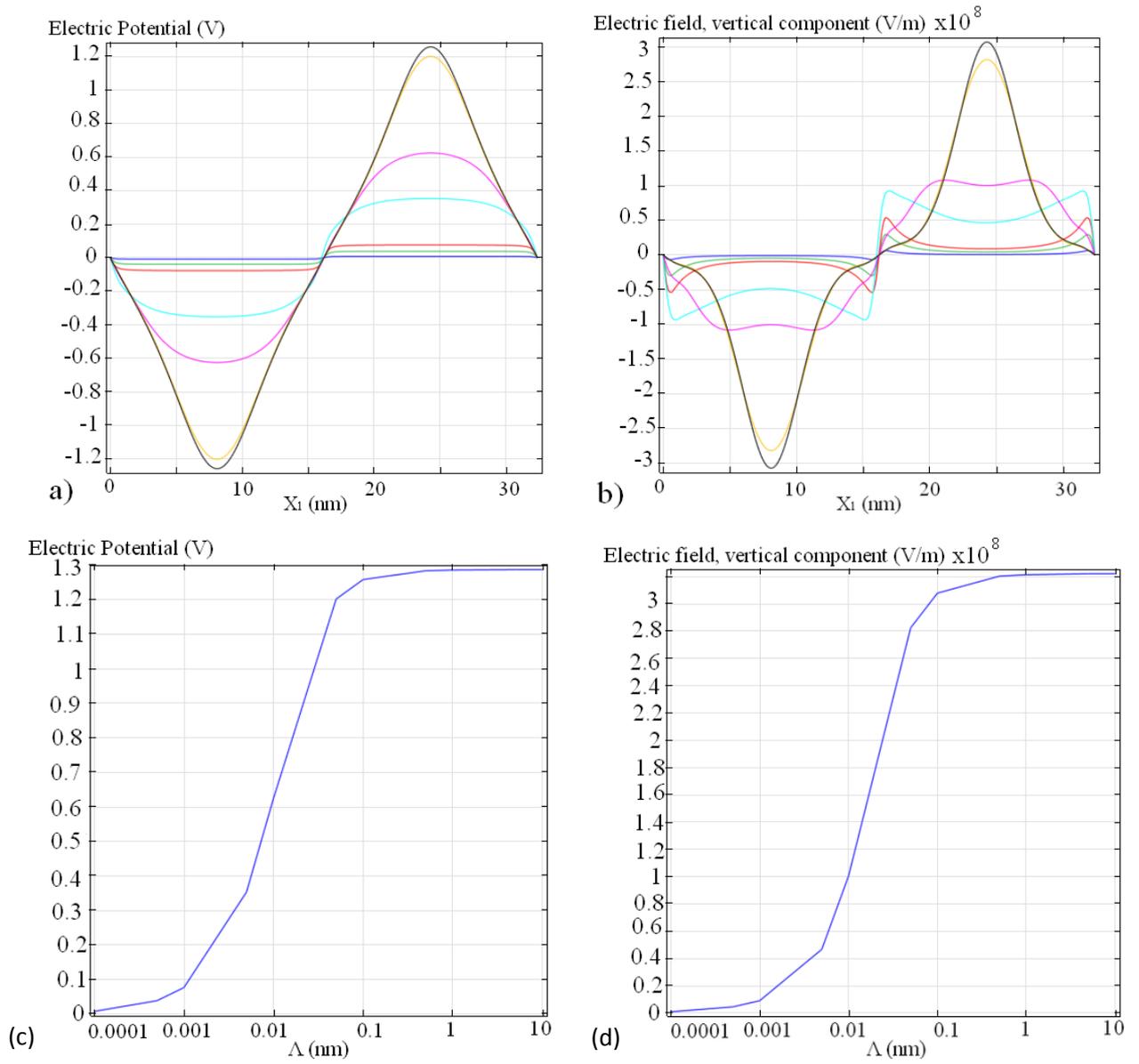

**FIG. S2**. The electric potential (a, c), the vertical component of the electric field (b, d) at the top surface dependencies of on the parameter of $\Lambda$. (a, b) Distribution along the surface at different $\Lambda$ (0.0001 nm - blue curve, 0.0005 - green, 0.001 - red, 0.005 - cyan, 0.01 - magenta, 0.05 - yellow, 0.1 - black); (c, d) the values in the middle of the domain.

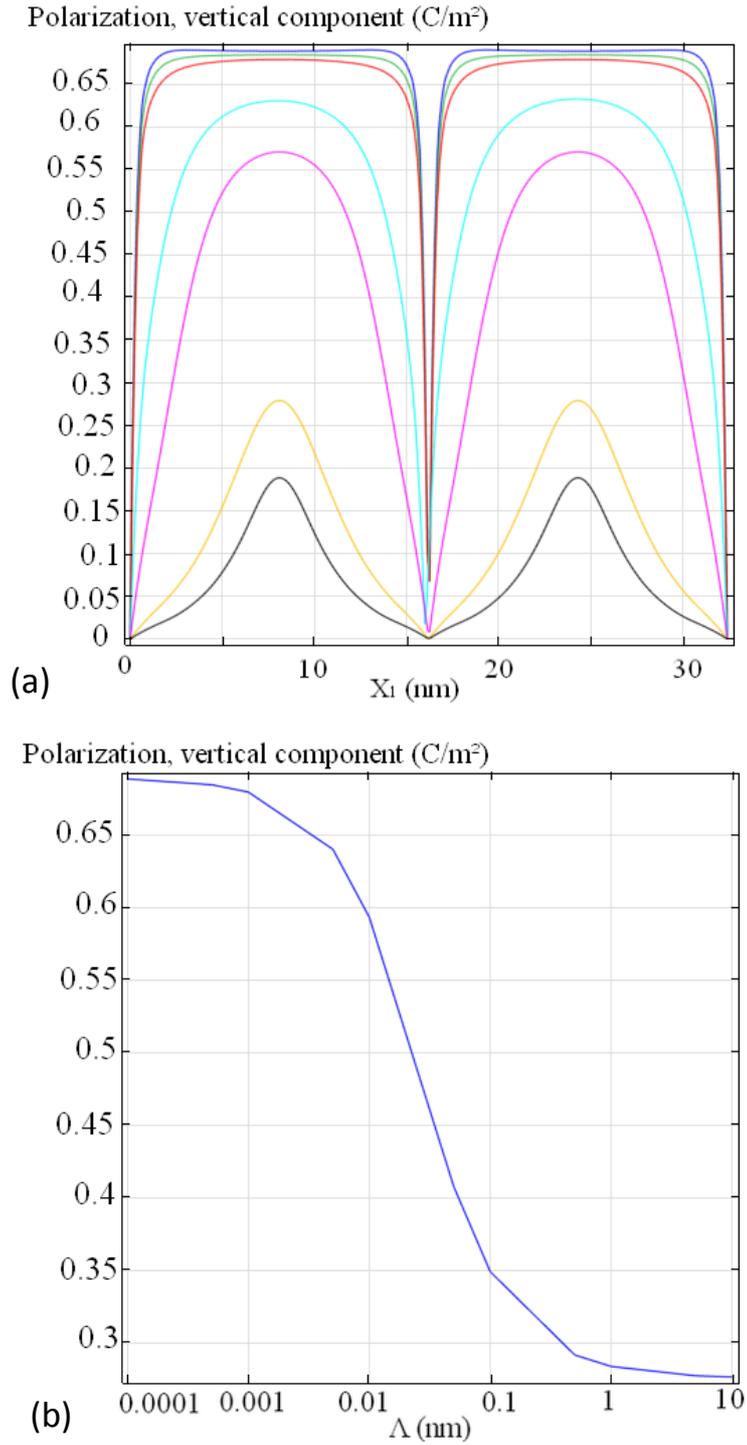

**FIG. S3**. (a) Distribution of the polarization along the surface at different Λ (0.0001 nm - blue curve, 0.0005 - green, 0.001 - red, 0.005 - cyan, 0.01 - magenta, 0.05 - yellow, 0.1 - black). (b) The dependence of the maximal polarization on the parameter of Λ. Polarization was calculated at a distance one unit cell from the film surface.

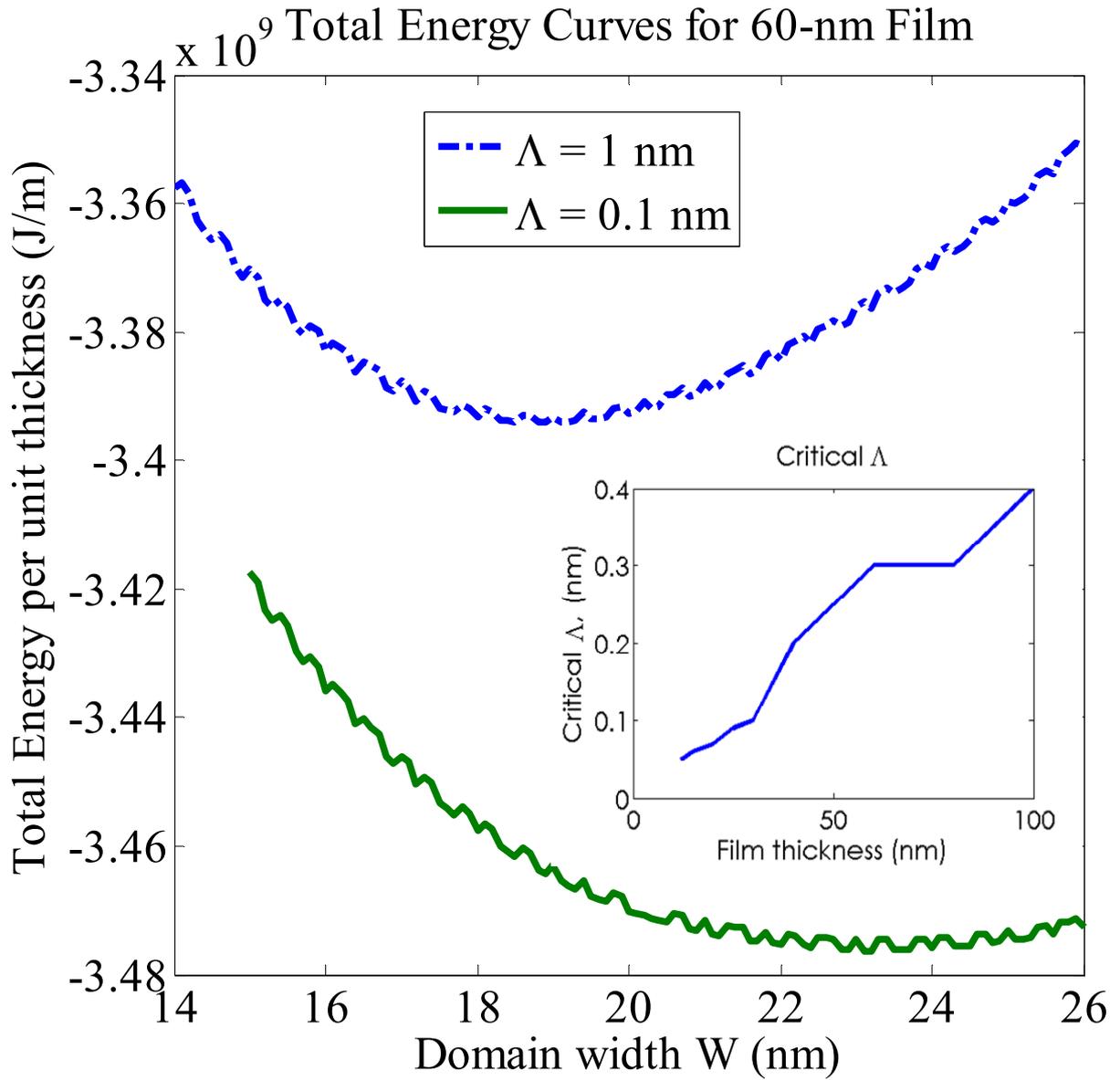

**FIG. S4**. Evaluation of the equilibrium domain width for 60-nm film through the minimum of total energy of the system at room temperature and Λ=1 nm (blue dash-dotted curve) and Λ=0.1 nm (green solid curve). It can be seen that the domain width with minimum energy for this film is estimated as 18.8 nm (Λ=1 nm) and 23.2 nm (Λ=0.1 nm).

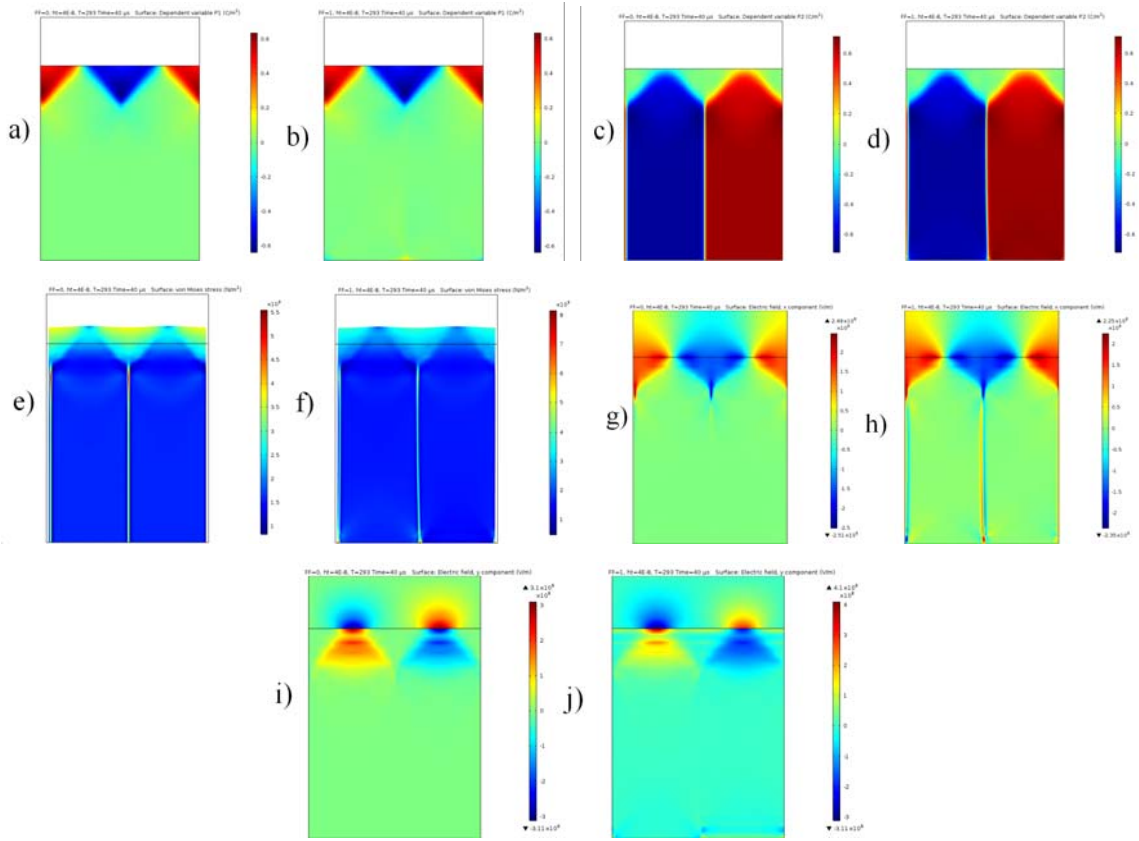

**FIG. S5.** Comparison between spatial distributions of horizontal (a, b) and vertical (c, d) spontaneous polarization components, elastic stress (e, f), horizontal (g, h) and vertical (i, j) electric field components in the 40-nm PT film without (a, c, e, g, i) and with (b, d, f, h, j) flexoelectric effect at the screening length $\Lambda$ = 0.1 nm. Flexoelectricity creates and stabilizes tiny closure nanodomains near the film-electrode interface with an increased mechanical stress in these areas.

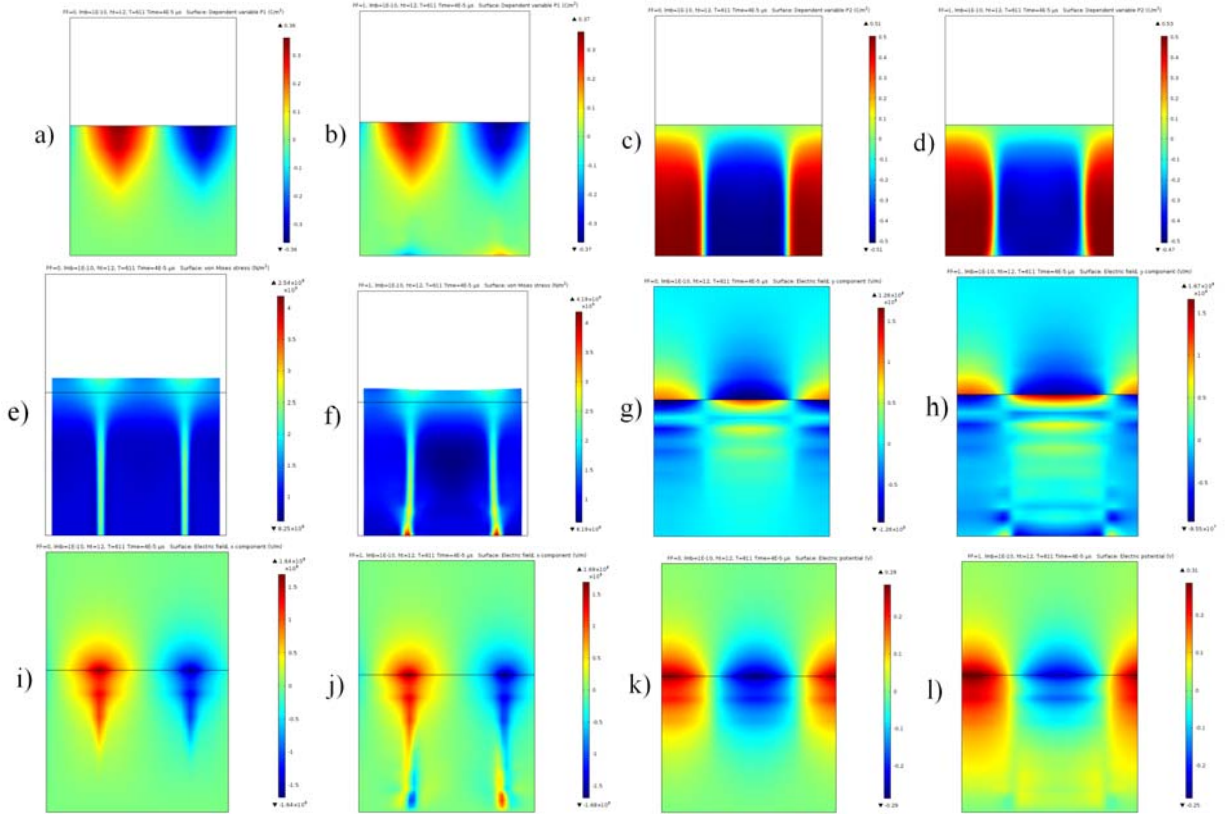

**FIG. S6.** Comparison between spatial distributions horizontal (a, b) and vertical (c, d) spontaneous polarization components, elastic stress (e, f), vertical (g, h) and horizontal (i, j) electric field components, and electrostatic potential (k, l) in the 12-nm PT film without (a, c, e, g, i, k) and with (b, d, f, h, j, l) flexoelectric effect at the screening length $\Lambda = 0.1$ nm and the temperature 611 K. Flexoelectricity creates and stabilizes tiny closure nanodomains near the film-electrode interface with an increased mechanical stress in these areas. Film thickness was chosen close to the critical value of the thickness induced phase transition for the ferroelectric film.

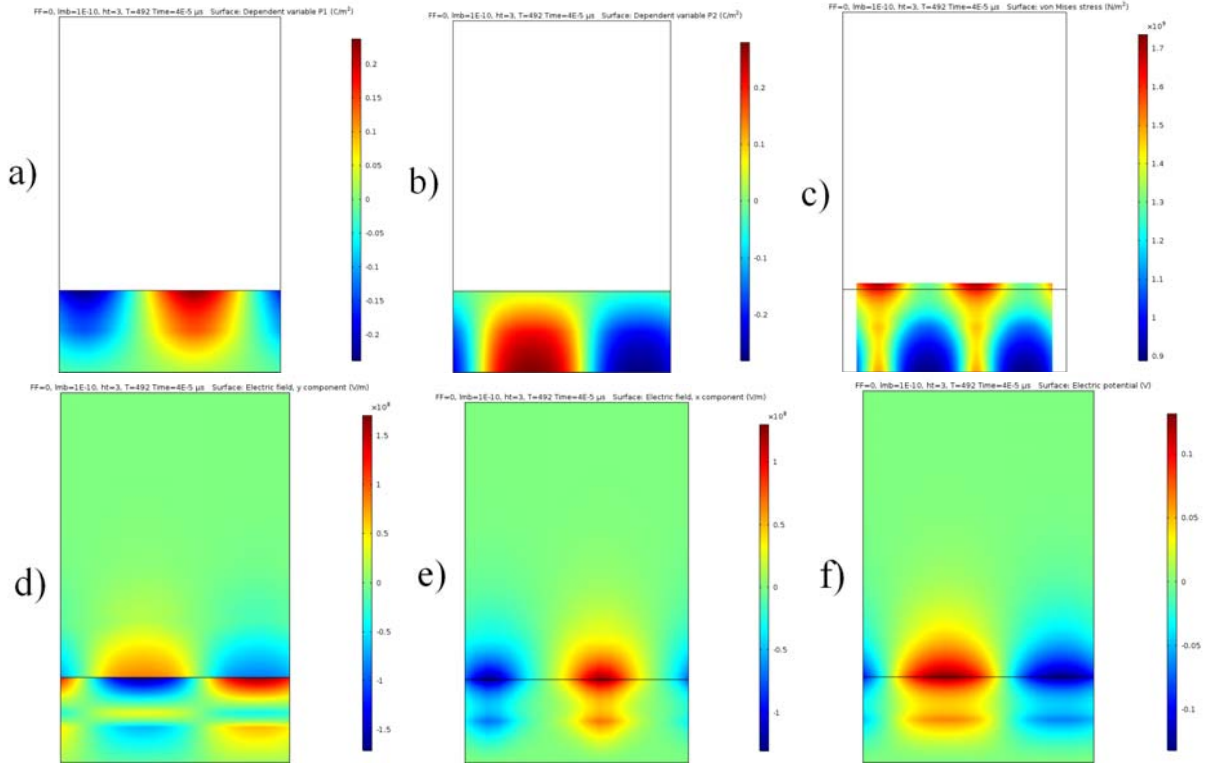

**FIG. S7.** Spatial distributions of horizontal and vertical polarization components P1 and P2 (a, b), mechanical stress (c), vertical and horizontal electric field components (d, e), and electrostatic potential (f) throughout a 3-nm thin film of PT at the temperature 492 K. Λ=0.1 nm, FA=0.

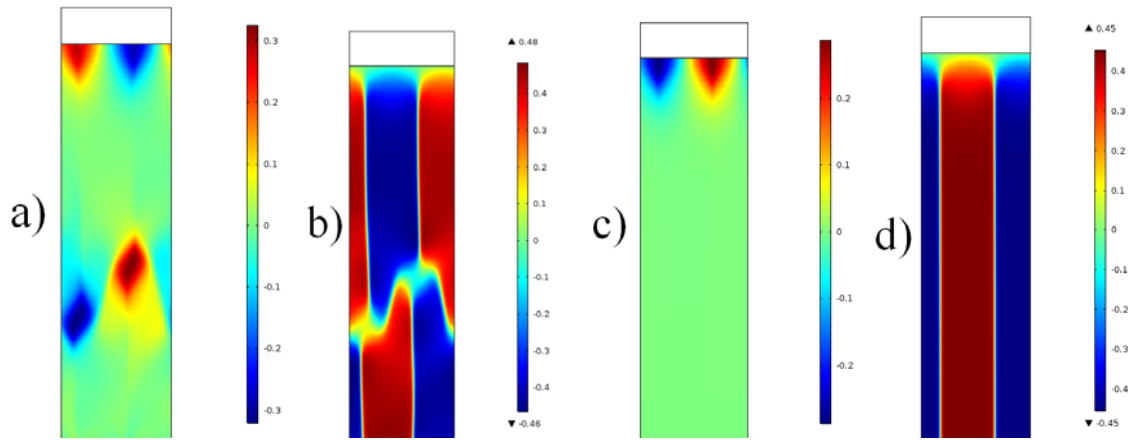

**FIG. S8.** Spatial distribution of the horizontal (a, c) and vertical (b, d) components of spontaneous polarization in the 110-nm-thick film with (a, b) and without (c, d) flexocoupling, featuring stabilized by flexoelectricity horizontal domains in the bulk with domain dimensions (≈10×20 nm) similar to those of polar nanoregions. Λ=0.1 nm.